\documentclass[aps,prl,preprint,superscriptaddress]{revtex4-1} 
\pdfpageattr{/Group <</S /Transparency /I true /CS /DeviceRGB>>}

\usepackage{siunitx}
\usepackage{graphicx}
\usepackage{epsfig}
\usepackage{bm}
\usepackage{amsmath}
\usepackage{amssymb}
\usepackage{wasysym}
\usepackage{epstopdf}
\usepackage{color}
\usepackage{xcolor}
\usepackage{epstopdf}
\usepackage{upgreek}
\usepackage{natbib}
\usepackage{arydshln}
\usepackage[T1]{fontenc}
\usepackage{subfig}
\usepackage[colorlinks=false, bookmarks=true]{hyperref}

\begin{document}


\title{Boiling regimes of impacting drops on a heated substrate under reduced pressure}



\author{Michiel A. J. van Limbeek}
\affiliation{Physics of Fluids Group, Mesa+ Institute, University of Twente, 7500 AE Enschede, The Netherlands}
\affiliation{Max Planck Institute for Dynamics and Self-Organization, 37077 G\"ottingen, Germany}
\author{Paul B. J. Hoefnagels}
\affiliation{Physics of Fluids Group, Mesa+ Institute, University of Twente, 7500 AE Enschede, The Netherlands}

\author{Minori Shirota}
\affiliation{Faculty of Science and Technology, Hirosaki University, 0368561, Aomori, Japan}

\author{Chao Sun}
\email{Corresponding author: chaosun@tsinghua.edu.cn (CS)}
\affiliation{Center for Combustion Energy and Department of Thermal Engineering, Tsinghua University, 100084 Beijing, China}
\affiliation{Physics of Fluids Group, Mesa+ Institute, University of Twente, 7500 AE Enschede, The Netherlands}
\affiliation{Max Planck Institute for Dynamics and Self-Organization, 37077 G\"ottingen, Germany}

\author{Detlef Lohse}
\email{Corresponding author: d.lohse@utwente.nl (DL}

\affiliation{Physics of Fluids Group, Mesa+ Institute, University of Twente, 7500 AE Enschede, The Netherlands}
\affiliation{Max Planck Institute for Dynamics and Self-Organization, 37077 G\"ottingen, Germany}



\date{\today}
\begin{abstract}We experimentally investigate the boiling behavior of impacting ethanol drops on a heated smooth sapphire substrate at pressures ranging from $P = 0.13$ bar to atmospheric pressure. We employ Frustrated Total Internal Reflection (FTIR) imaging to study the wetting dynamics of the contact between the drop and the substrate. The spreading drop can be in full contact (contact boiling), it can partially touch (transition boiling) or the drop can be fully levitated (Leidenfrost boiling). We show that the temperature of the boundary between contact and transition boiling shows at most a weak dependency on the impact velocity, but a significant decrease with decreasing ambient gas pressure. A striking correspondence is found between the temperature of this boundary and the static Leidenfrost temperature for all pressures. We therefore conclude that both phenomena share the same mechanism, and are dominated by the dynamics taken place at the contact line.\\
On the other hand, the boundary between transition boiling and Leidenfrost boiling, i.e. the dynamic Leidenfrost temperature, increases for increasing impact velocity for all ambient gas pressures. Moreover, the dynamic Leidenfrost temperature coincides for pressures between $P = \SI{0.13}{}$ and $P= \SI{0.54}{}$ bar, whereas for atmospheric pressure the dynamic Leidenfrost temperature is slightly elevated. This indicates that the dynamic Leidenfrost temperature is at most weakly dependent on the enhanced evaporation by the lower saturation temperature of the liquid.
\end{abstract}

\pacs{ *43.25.Yw, *43.25.Yw, *43.35.Ei}
\keywords{Leidenfrost effect, Nucleate boiling,  TIR-imaging, Droplet heat transfer}

\maketitle


\section{Introduction}
\footnotetext{\dag~Electronic Supplementary Information (ESI) available: [details of any supplementary information available should be included here]. See DOI: 10.1039/b000000x/}

\footnotetext{$^{\ast}$~ \textit{corresponding authors: c.sun@utwente.nl and d.lohse@utwente.nl}}
\footnotetext{\textit{$^{a}$~Physics of Fluids, University of Twente, P. O. Box 217, 7500 AE Enschede, The Netherlands}}
\footnotetext{\textit{$^{b}$~Center for Combustion Energy and Department of Thermal Engineering, Tsinghua University, 100084 Beijing, China }}
\footnotetext{\textit{$^{c}$~Max Planck Institute for Dynamics and Self-Organization, 37077 G\"{o}ttingen, Germany}}




Drop impact on heated surfaces has enjoyed a lot of research attention over the recent years \cite{liang2017,Josserand2016} due to its numerous applications, such as in spray cooling, burn-out phenomena and fuel injection \cite{Kim2007,Moreira2010}. In these processes it is desirable to have a high heat transfer rate between the solid and the liquid which is achieved by a large contact area between the liquid and the substrate. When a substrate is heated above a certain temperature, the drop will hover above the substrate on its own vapour. The low thermal conductivity of the vapour layer insulates the liquid from the substrate, hence the heat transfer from the substrate to the drop is drastically reduced. As a result, the lifetime of a deposited drop increases significantly. The lowest substrate temperature $T_\mathrm{sur}$ at which this phenomenon occurs is referred to as the Leidenfrost temperature $T_\mathrm{L}$ \cite{Leidenfrost1756,Biance2003,Quere2013}.
Under reduced ambient pressures the saturation temperature $T_\mathrm{sat}$ decreases as well, as described by the Antoine equation 
\begin{equation}
T_\mathrm{sat}=\frac{b}{a-\log_{10} \left( \frac{P}{P_0} \right)}-c
\end{equation}
that is derived from the Clausius-Clapeyron equation. Orejon et al. \cite{Orejon2014} suggested an simplified form of this equation to describe $T_\mathrm{L,static}$ as a function of pressure:

\begin{equation}
	T_\mathrm{L,static} = \frac{1}{A - B \cdot \log_{10} \left( \frac{P}{P_0} \right)},
	\label{eq:T_L}
\end{equation}

with $P_0 = \SI{1}{\bar}$. Using this relation Orejon et al. \cite{Orejon2014} were able to fit $T_\mathrm{L,static}(P)$ to their experimental results for a wide range of liquids and substrates. This raises the question how well this relation applies to the dynamic Leidenfrost temperature, i.e. the case when the drop is forced onto the heated substrate by an inital downward motion. \\

\begin{figure}[h!]
\centering
\def\svgwidth{0.75\textwidth}
\includegraphics[scale=1]{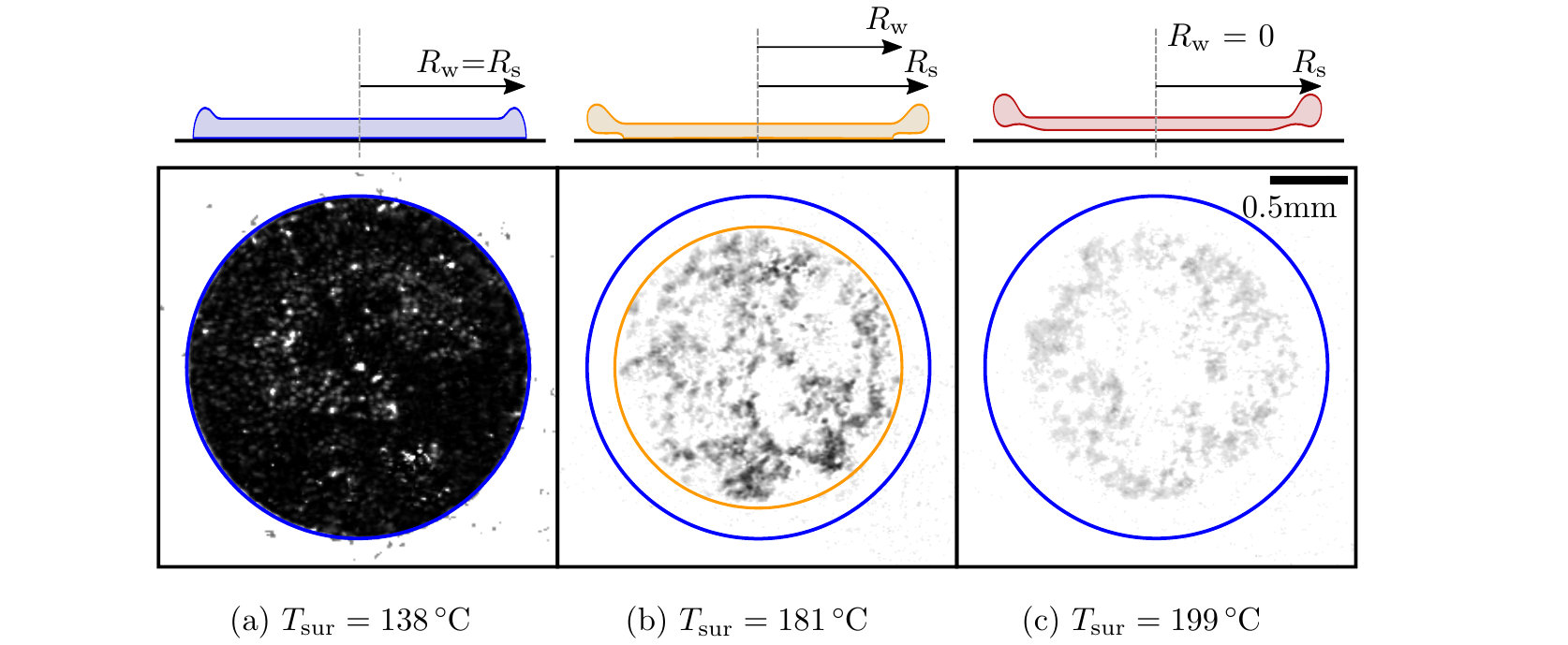}
\caption{FTIR images of ethanol drops at $t =  \SI{0.21}{\milli \second}$ after impacting a sapphire prism with impact velocity $U = \SI[per-mode=symbol]{1.0}{\meter \per \second}$ at atmospheric pressure. For a substrate temperature of $T_\mathrm{sur} = \SI{138}{\degreeCelsius}$ the spreading drop fully wets the surface and is in contact boiling. At $\SI{181}{\degreeCelsius}$ the drop partially wets the surface and identified as transition boiling. At $\SI{199}{\degreeCelsius}$ the drop is only visible in gray values: it is fully levitated and hence in the Leidenfrost boiling state. The blue circles correspond to the radius of the wetted area of a drop in the contact boiling state.}
\label{fig:Boilingstates} 
\end{figure}

\begin{figure}[h!]
\centering
\includegraphics[]{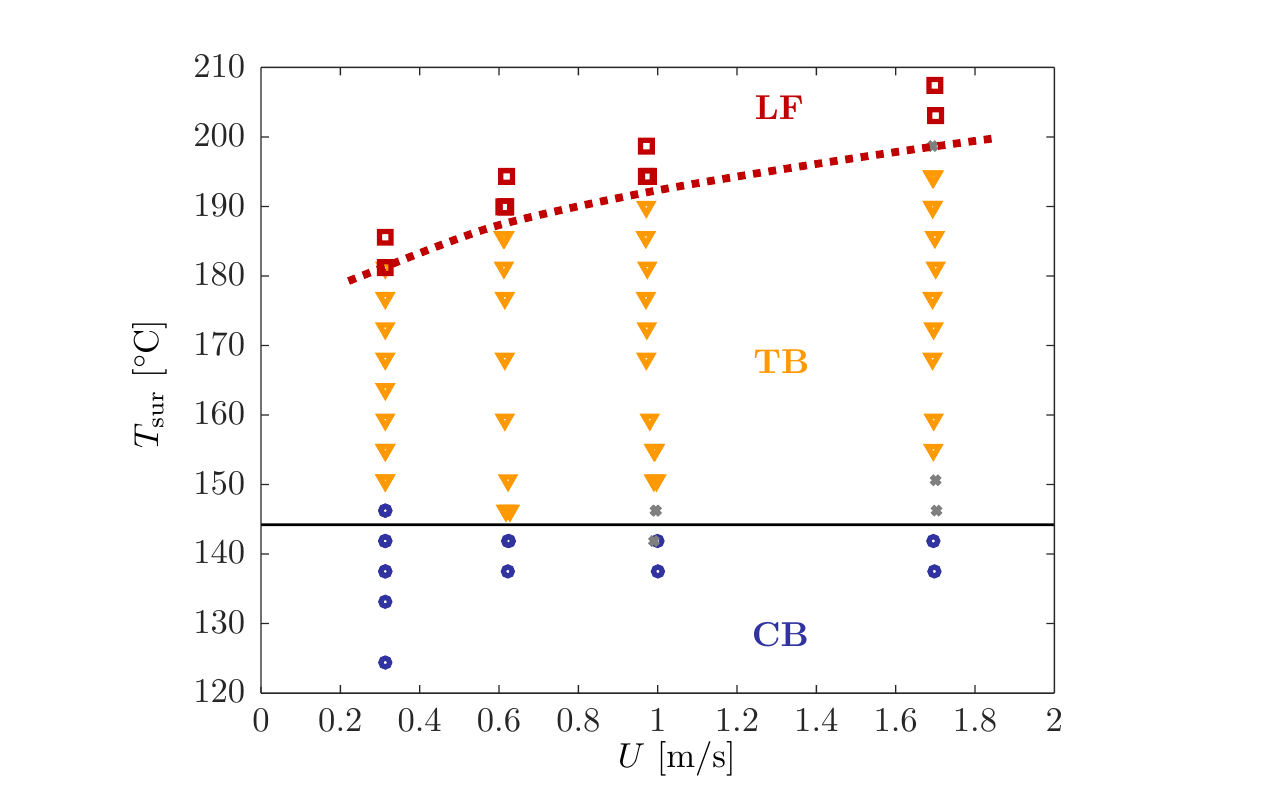}
\caption{Phase diagram for ethanol drops impacting on sapphire substrate at atmospheric gas pressure. The blue diamonds correspond to contact boiling, yellow triangles to transition boiling and red squares to the Leidenfrost state. The grey crosses indicate measurements for which the boiling state is hard to identify. The red dotted line is a guide to the eye for the TB-LF boundary: the dynamic Leidenfrost temperature. The horizontal black line indicates the static Leidenfrost temperature on a silicon substrate for comparison.}
\label{fig:PhaseDiagram_Patm}
\end{figure}

For drops impacting a heated surface at atmospheric pressure it is shown that $T_\mathrm{L}$ increases with increasing impact velocity \cite{Tran2012,Chandra1991,Wachters1966}. Prior to impact a high pressure region is formed between the drop and the substrate, a result of the viscosity of the escaping air \cite{Josserand2016}. This causes a deceleration of the drop and the formation of a dimple in the drop before touchdown \cite{Eggers2010,Bouwhuis2012}. Touchdown of the drop on the substrate can be prevented completely in the case of a heated substrate, because the escaping air is replenished by the evaporation of the drop \cite{Tran2012,Chandra1991,Shirota2016}. A more sytimatical study was performed by Shirota et al. \cite{Shirota2016}, where the boiling behaviour was charaterized into three boiling regimes: contact boiling (CB), transition boiling (TB) and Leidenfrost boiling (LF), depending on the relative behaviour between the wetting and spreading radius. An example is given in \autoref{fig:Boilingstates}, where with increasing plate temperature the wetting radius diminishes and finaly vanishes. The temperature at which a change in boiling behaviour occurs between contact and transition boiling will be referred to as the CB-TB boundary. Similarly, the TB-LF boundary refers to the temperature at which drops enter the Leidenfrost state from the transition boiling regime. A phase diagram for ethanol drops with different impact velocities $U$ and surface temperatures $T_\mathrm{sur}$ at atmospheric pressure is shown in \autoref{fig:PhaseDiagram_Patm}, reproduced in the current study as a reference to and in agrement with previous results \cite{Shirota2016,vanLimbeek2016}. \\

Since it is known that the ambient gas pressure has a great influence on impact dynamics \cite{Xu2005,Xu2007a,Xu2007b,Mandre2009,Mani2010,Duchemin2011,Kolinski2012,Latka2012,Stevens2014,Riboux2014}, it is interesting to study how heating the substrate affects the touchdown under reduced pressure conditions. A major consequence of lowering the ambient gas pressure in the unheated case is the suppression of splash formation \cite{Xu2005}. This phenomenon has been studied extensively both experimentally \cite{Xu2005,Xu2007a,Xu2007b, Duchemin2011,Kolinski2012,Latka2012,Stevens2014} and numerically \cite{Mandre2009,Mani2010}, where a crucial role is found for both the wetting of the substrate and the air-liquid interaction \cite{Riboux2014}. However since the vapor generation is affected by the change in saturation temperature, it is far from obvious how such a system will behave under various  ambient gas pressures. The goal of this study is to investigate the role of the pressure on  the wetting dynamics and boiling behaviors.

\section{Experimental aspects}
\begin{figure}[ht!]
\centering
\includegraphics[scale=1]{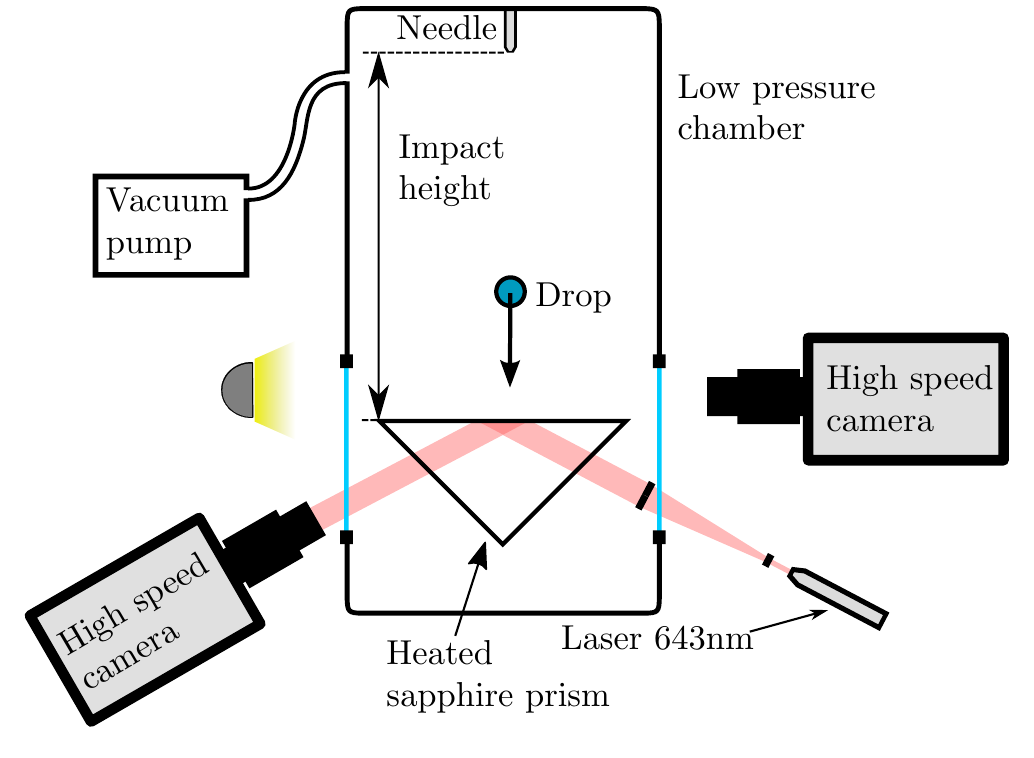}
\caption{Schematic representation of the set-up used in our experiments. In addition to the side-view camera, we study the wetting of the drop by a bottom-view camera, using frustrated total internal reflection imaging. The sapphire prism was heated to set the desired surface temperature $T_\mathrm{sur}$.}
\label{fig:Setup}
\end{figure}

To study the impact behaviour of ethanol drops on a heated surface at various pressures, an FTIR (Frustrated Total Internal Reflection) setup is used as shown in \autoref{fig:Setup}. Details of the setup and methods are described before, \cite{Shirota2016,vanLimbeek2016,Shirota2017}, for self-consistency, here we give the essential aspects. A laser beam (wavelength $\lambda = \SI{643}{\nano \meter}$, p-polarized) is widened by a beam expander and illuminates the sapphire-air surface. The  angle of the incident light is larger than the critical angle to reflect the laser beam. The reflected light is recorded with a high speed camera (Photron SA-X2) with a frame rate of at least $\SI{20000}{}$ fps, and the camera is focused on the top interface of the prism. When a drop is in contact with the surface, a fraction of the laser beam can be transmitted across the interface, resulting in the detection of a dark spot by the camera. Intermediate intensities of the reflected light appear when the base of the drop is just above the surface within the distance of the decay length of the evanescent wave, typically in the order of hundreds of $\SI{}{nm}$. A sapphire substrate is chosen because of its transparency, its smoothness (surface roughness of $\SI{10}{\nano \meter}$) and its relatively high thermal conductivity of $k_\mathrm{sur} = \SI[per-mode=symbol]{32}{\watt \per \meter \per \K}$. The sapphire prism is positioned in an aluminium holder that is kept at a constant temperature using a proportional-integral-derivative controller. The temperature of the sapphire surface $T_\mathrm{sur}$  is calibrated by using a surface probe (N-141K, Anritsu) as well as a Pt-100 sensor for various ambient pressures and heater set points.

The needle and heated surface are situated in a low pressure chamber. The surrounding pressure ranging from $P = \SI{0.13}{\bar}$ to atmospheric pressure, is controlled by a pump and a pressure sensor (JUMO dTrans p30). The chamber has a volume of about $\SI{30}{\liter}$, so the evaporated ethanol drops has a negligible effect on the gas composition. The needle is surrounded by a brass block that is maintained at a constant temperature of $\SI{5}\degreeCelsius$ to ensure that the ethanol is well below the saturation temperature at all pressures. 
Single drops of diameter $D_0 = \SI{2.6}{\milli \meter}$ are created from the tip of the needle and fall down when the gravitational force on the drop overcomes the surface tension. The impact velocity $U$ on the substrate is varied between $U = \SI[per-mode=symbol]{0.3}{\meter \per \second}$ and $\SI[per-mode=symbol]{1.7}{\meter \per \second}$ by varying the height of the needle. 
$D_0$ and $U$ are measured by tracking the drop $\SI{4}{\milli \meter}$ before impact using a secondary high speed camera (Photron SA1.1) at $\SI{20000}{}$ fps.
The light sources and cameras are placed outside of the low pressure chamber and the images are recorded through a glass window in the chamber.

\section{Results and discussion}
\subsection{Pressure effect on static Leidenfrost effect}
\begin{figure}[]
\centering
\includegraphics[scale=1]{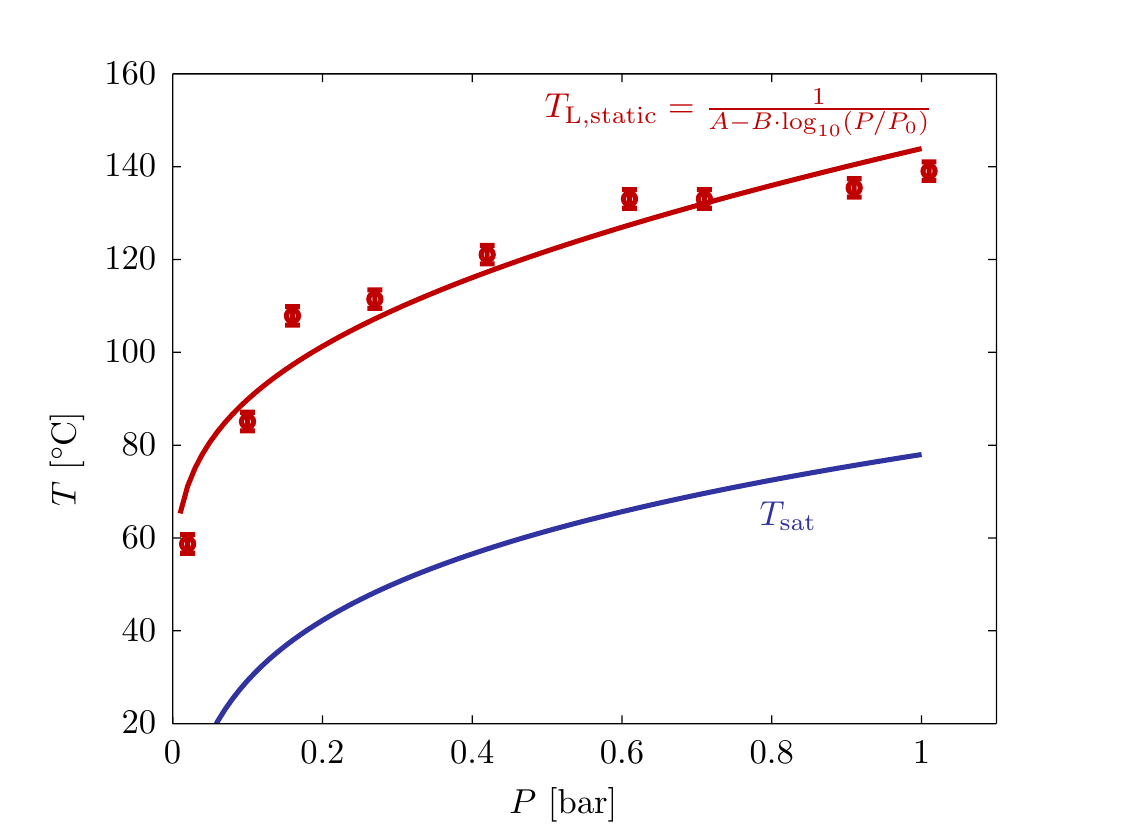}
\caption{The static Leidenfrost temperature $T_\mathrm{L}$ of ethanol drops on silicon changes with ambient pressure, as shown in red circles. Following Orejon et al. \cite{Orejon2014}, a logarithmic function (\autoref{eq:T_L}) is fitted through these points, with $A = \num{0.0069+-0.0005}$  and $B = \num{0.004+-0.001}$. The dependence of the saturation temperature on the pressure of ethanol is shown as the blue solid line.}
\label{fig:VaporPressure}
\end{figure}

Since the change in boiling behavior from contact to transition boiling correlates strongly with the static Leidenfrost temperature \cite{Shirota2016,vanLimbeek2016}, we first perform experiments for ethanol to obtain $T_\mathrm{L}$ for reduced pressure conditions. Orejon et al. \cite{Orejon2014} found the static Leidenfrost temperature $T_\mathrm{L,static}$  decreasing with pressure for various liquids, however no results are available for ethanol, the liquid of interest in the current study. The pressure dependency of $T_\mathrm{L}=T_\mathrm{L}(P)$ is measured here on a silicon substrate of \SI{5}{\milli \meter} thickness, since this can be considered to be isothermal \cite{Baumeister1973, vanLimbeek2017b} during the evaporation of the drop, mainly due to the high thermal conductivity of silicon ($k_s = \SI{148}{\watt \per \meter \per \kelvin}$ at $T = \SI{300}{\kelvin}$). The results are shown in \autoref{fig:VaporPressure}, with a best fit of \autoref{eq:T_L} giving fitting parameters $A = \num{0.0069+-0.0005}$  and $B = \num{0.004+-0.001}$ and $P_0$ the (reference) ambient pressure.
This fitting shows a similar trend as the saturation temperature $T_\mathrm{sat}$ of ethanol. It is not surprising that $T_\mathrm{L}$ decreases with decreasing ambient pressure, as then the escaping vapor under the drop has to do less work against the smaller ambient pressure. Similarly, the evaporation is easier as the saturation temperature decreases, resulting in a comparable temperature difference between $T_\mathrm{L}(P)$ and $T_\mathrm{sat}(P)$. 
\subsection{Phase diagram under reduced pressure}
\label{sec:amb}
\begin{figure}[]
\centering
\def\svgwidth{0.75\textwidth}
\includegraphics[scale=1]{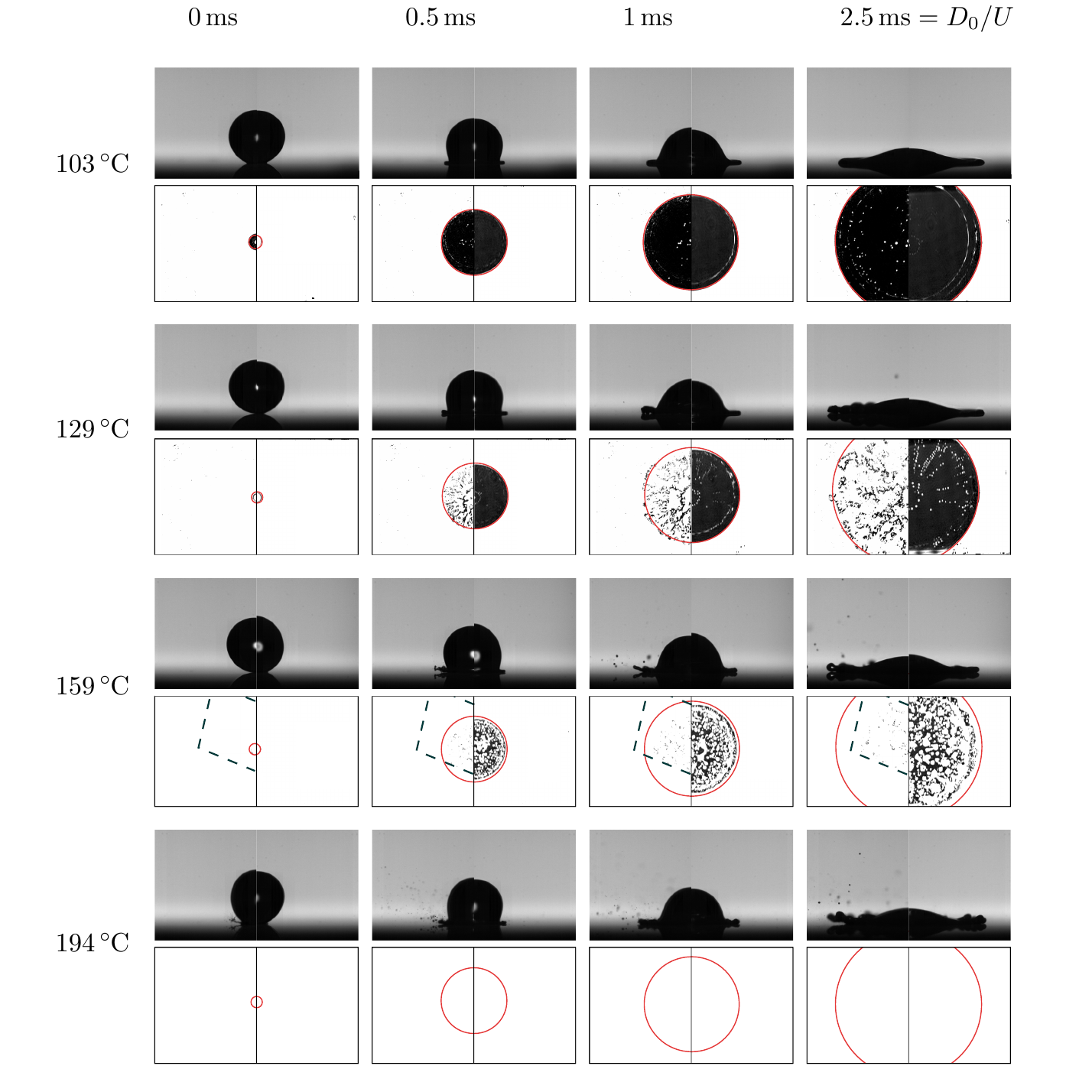}
\caption{Comparison of ethanol drops impacting a heated sapphire prism under an ambient pressure of 0.29 and \SI{1}{\bar} (resp. left and right half of every subfigure ). In addition to the side view observations, the FTIR views are displayed, revealing a clear distinction between wet areas (black) and air/vapor regions (white). The spreading radius as found by the side view observations is included in red as a reference, revealing the the changes in boiling state with increasing temperature and reduction in pressure. The dashed line for the $T_\mathrm{sur}=\SI{159}{\celsius}$ case indicates the field of view for this particular measurent.}
\label{fig:ResImp}
\end{figure}

The boiling behavior of the impacting ethanol drops on the heated surface is studied in detail using the FTIR images, enabling us to study the wetting behavior during impact. A selection of the impacts are displayed in \autoref{fig:ResImp}, where for four different plate temperatures a series of snapshots is presented. The subfigures are composed of an impact in ambient conditions (right half) and at a reduced pressure of \SI{0.29}{\bar}, where both the side-view and FTIR-recordings are shown. It is clearly visible form the recordings at $T_\mathrm{sur}=\SI{103}{\celsius}$ that for both pressures the wetted (black) area of the FTIR recordings  corresponds with the spreading behavior as observed by the side view camera. This radius is imposed for all FTIR recordings as a reference for the wetting behavior. From these results, two trends can directly be deduced: Increasing the initial temperature of the substrate suppresses the wetting of the plate and secondly, reducing the ambient pressure yields the same trend.

As suggested by  Shirota et al. the impacts can be categorized into three boiling regimes: contact boiling (CB), transition boiling (TB) and Leidenfrost boiling (LF) \cite{Shirota2016}. At sufficiently low substrate temperatures $T_\mathrm{sur}$, the wetting front of the drop has the same velocity as that of a drop impacting an unheated surface \cite{Shirota2016}. This is identified as the contact boiling state, see the $T_\mathrm{sur}=\SI{103}{\celsius}$ recordings of \autoref{fig:ResImp}. 
At certain $T_\mathrm{sur}$ this is no longer observed: the spreading radius $R_\mathrm{s}$ of the drop exceeds the wetting radius $R_\mathrm{w}$ of the liquid on the surface. Examples are shown in  \autoref{fig:ResImp} at  $T_\mathrm{sur}=\SI{129}{\celsius}$ at \SI{0.29}{bar} and $T_\mathrm{sur}=\SI{159}{\celsius}$ at abient pressure. The edge of the drop is levitated above the surface, called the lamella, and this behavior is categorized as transition boiling \cite{Shirota2016}.
At even higher values of $T_\mathrm{sur}$ the impacting drop does not wet the surface at all, hence it is in the Leidenfrost state, see for instance the recordings at $T_\mathrm{sur}=\SI{194}{\celsius}$. 
The temperature at which a change in boiling behavior occurs between contact and transition boiling will be referred to as the CB-TB boundary. Similarly, the TB-LF boundary refers to the temperature at which drops enter the Leidenfrost state from the transition boiling regime.

The resulting  phase diagram for ethanol drops with different impact velocities $U$ and surface temperatures $T_\mathrm{sur}$ at atmospheric pressure was introduced earlier in \autoref{fig:PhaseDiagram_Patm}. The temperature for the CB-TB boundary is at most weakly dependent of $U$, while the temperature of the TB-LF boundary, i.e. the dynamic Leidenfrost temperature, increases for higher $U$. This is in agreement with previous results for different liquids \cite{Shirota2016,vanLimbeek2016,vanLimbeek2017c}. \\

\begin{figure}[]
\centering
\includegraphics[scale=1]{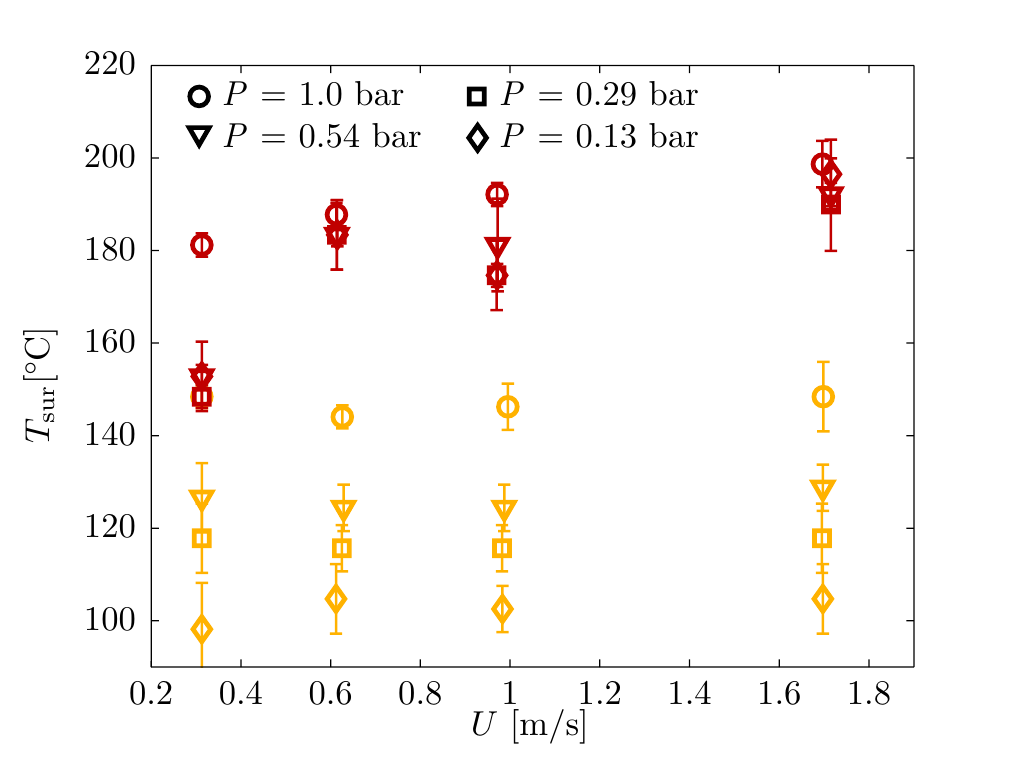}
\includegraphics[scale=1]{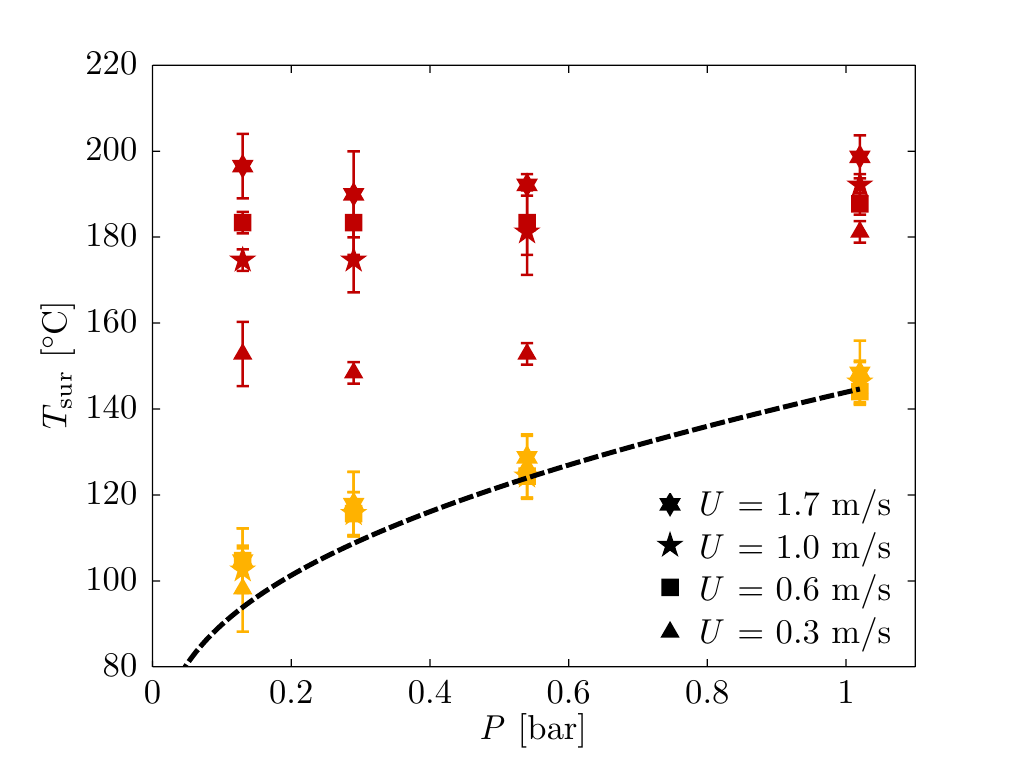}
\caption{The CB-TB boundaries for different pressures and impact velocities are shown in yellow and the dynamic Leidenfrost temperatures are shown in red. Both plots show the same data, but they are represented in a different to show different effects more clearly. In (a) the boundaries are plotted as a function of impact velocity, in (b) the boundaries are plotted as a function of ambient pressure.  It can be seen that the CB-TB boundary is almost independent of impact velocity and is in close agreement with $T_\mathrm{L,static}(P)$ indicated with the black dotted line in (b). The TB-LF boundary increases for most measurement with impact velocity and change slightly with pressure.}
\label{fig:PhaseDiagram_Pall}
\end{figure}

\subsubsection{Pressure effect on boundary between the contact and transition boiling regime}
For all ambient pressures the CB-TB boundaries are within the measurement error independent of impact speed, see \autoref{fig:PhaseDiagram_Pall}. This is in correspondence with the trend found for measurements at atmospheric pressure \cite{Shirota2016,vanLimbeek2016}. However, the temperature of the CB-TB boundary lowers with decreasing ambient gas pressure. When this boundary is compared to $T_\mathrm{L,static}(P)$ the agreement  is striking, suggesting that the latter is determined by the same physical mechanisms as $T_\mathrm{L}(P)$.

\subsubsection{Pressure effect on the dynamic Leidenfrost temperature}
An interesting behavior of the dynamic Leidenfrost temperature is found when varying the ambient gas pressure: we find a collapse for impacts at ambient gas pressures between $P = \SI{0.13}{}$ and $P = \SI{0.54}{\bar}$, as shown in \autoref{fig:PhaseDiagram_Pall}. When comparing the dynamic Leidenfrost temperature under reduced pressures with our results at atmospheric conditions we find a slightly lower temperature. On the other hand, the dynamic Leidenfrost temperature lowers with decreasing impact velocity for all ambient gas pressures.

It highlights that the impact dynamics are dominated by the high pressure developing in the drop \cite{Wagner1932,Howison1991,Zhao1993,Philippi2016}, rather than the pressure in far field. The complexity of the dynamic Leidenfrost problem is now further increased due to the compressibility of the gas in reduced pressure conditions \cite{Mani2010}. Although an enhancement of the evaporation is expected as a result of the lower saturation temperature in reduced pressure conditions, our observations indicate that this does not play a dominant role for the dynamic Leidenfrost temperature. We conclude that the ambient gas pressure has minor influence on the local pressure underneath the drop and thus on the vapour pressure and evaporation rate. 

Numerical work showed that at lower ambient gas pressures the shape of the impacting drop changes and the neck widens \cite{Mani2010}. Since a larger area is close to the wall, an enhanced vapor generation might be expected, compensating the lower content of air at reduced pressure conditions, but the current work does not allow us to draw firm conclusions on this matter.


\section{Conclusions}
In this study we identified the boiling regimes for ethanol drops on a sapphire substrate for different impact velocities and ambient gas pressures. High speed FTIR imaging is used to observe the wetting of the impacting drops. 

The temperature of the boundary between contact boiling and transition boiling systematically  decreases when reducing the ambient gas pressure. It showed excellent agreement with the Leidenfrost temperature found here for static drops, from which we conclude that they share the same underlying mechanism. This boundary remained approximately  independent of impact velocity, in agreement with previous studies.

On the other hand, we found a collapse of the dynamic Leidenfrost temperature at reduced pressure conditions, slightly below the dynamic Leidenfrost temperature obtained in atmospheric conditions. 

Combining these two observations reveals that the temperature range where transition boiling is found strongly increases with decreasing pressure. It would be interesting to extend this study to elevated pressures to investigate how this trend develops above atmospheric conditions. One might find a collapse of both boundaries, resulting in the disappearance of the transition boiling regime.

\section*{Acknowledgements}
This work was supported by an ERC-Advanced Grant.

\bibliography{biblio} 
\end{document}